\documentclass[12pt]{article}
\usepackage{amssymb,latexsym,amsfonts,amsmath,amsthm}
\usepackage[dvips]{graphicx}
\usepackage[margin=3cm]{geometry}

\newcommand{\Cx}{\mathbb{C}}

\newcommand{\Ir}{\mathbb{Z}}

\newcommand{\Rl}{\mathbb{R}}

\newcommand{\E}{\mathbb{E}}

\renewcommand{\H}{\mathcal{H}}
\newcommand{\A}{\mathcal{A}}
\newcommand{\B}{\mathcal{B}}
\newcommand{\F}{\mathcal{F}}

\def\idty{{\mathchoice {\mathrm{1\mskip-4mu l}} {\mathrm{1\mskip-4mu l}} %
{\mathrm{1\mskip-4.5mu l}} {\mathrm{1\mskip-5mu l}}}}

\DeclareMathOperator{\Tr}{Tr}

\newcommand{\ket}[1]{\left\vert #1\right\rangle}

\newcommand{\spec}{\operatorname{spec}}

\newtheorem{theorem}{Theorem}[section]

\newtheorem{lemma}[theorem]{Lemma}

\newtheorem{conjecture}[theorem]{Conjecture}
\newtheorem{assumption}[theorem]{Assumption}

\renewcommand{\theequation}{\thesection.\arabic{equation}}
\newcounter{saveeqn}
\newcommand{\alpheqn}{\setcounter{saveeqn}{\value{equation}}%
  \stepcounter{saveeqn}\setcounter{equation}{0}%
  \renewcommand{\theequation}{%
  \mbox{\arabic{section}.\arabic{saveeqn}-\alph{equation}}}}
\newcommand{\reseteqn}{\setcounter{equation}{\value{saveeqn}}%
\renewcommand{\theequation}{\mbox{\arabic{section}.\arabic{equation}}}}
\newcommand{\eq}[1]{(\ref{#1})}

\newcommand{\be}{\begin{equation}}
\newcommand{\ee}{\end{equation}}
\newcommand{\bea}{\begin{eqnarray}}
\newcommand{\eea}{\end{eqnarray}}
\newcommand{\beann}{\begin{eqnarray*}}
\newcommand{\eeann}{\end{eqnarray*}}

\newcommand{\dom}{\mathop{\rm Dom}}

\begin{document}

\title{Quantum Spin Systems after DLS1978}

\author{Bruno Nachtergaele\\[10pt]
{Department of Mathematics}\\
{University of California, Davis}\\
{One Shields Avenue}\\ 
{Davis, CA 95616-8366, USA}\\
\texttt{bxn@math.ucdavis.edu}
}

\date{December 30, 2005}
\maketitle

\begin{abstract} 
In their 1978 paper, Dyson, Lieb, and Simon (DLS) proved the existence
of N\'eel order at positive temperature for the spin-$S$ Heisenberg
antiferromagnet on the  $d$-dimensional hypercubic lattice when either
$S\geq 1$ and $d\geq 3$ or $S=1/2$ and $d$ is sufficiently large. This
was the first proof of spontaneous breaking of a continuous symmetry in
a quantum model at finite temperature. Since then the ideas of DLS have
been extended and adapted to a  variety of other problems. In this paper
I will present an overview of the most important developments in the study
of the Heisenberg model and related quantum lattice systems since 1978,
including but not restricted to those directly related to the paper by
DLS.
\end{abstract}


\renewcommand{\thefootnote}{}
\footnotetext{Copyright \copyright\ 2006 Bruno Nachtergaele. This article may be
reproduced in its entirety for non-commercial purposes.}

\tableofcontents

\section{Introduction}

The purpose of this article is to review a number of important developments
in the mathematical physics of quantum spin systems that have occured since 
Barry Simon first got involved with the subject. A comprehensive account
of such developments clearly would not fit in the space alotted here. 
Therefore, I have selected seven topics
that either have a relation to the seminal paper by Dyson, Lieb and Simon
\cite{dyson1978} or that for some other reason I believe interesting and
important enough to deserve to be discussed in some detail here. In Section
\ref{sec:other} I list a number of works I regretfully had to leave out.

There is one development that stands out by its absence: a proof of
long-range order in the Heisenberg ferromagnet at positive temperature in 
dimensions $\geq 3$ has eluded the valiant attempts of many. One can trace
of some of these attempts in the literature \cite{conlon1990,toth1993,federbush2004},
and the authors of \cite{dyson1978} also comment on a erroneous proof they thought 
they had,
but certainly most failed approaches have not left a written record. It is not a hopeless
problem, in my opinion. It is much like other open phase transition problems such as hard-core
bosons at finite density and temperature in $\Rl^3$, or the $XY$-model in a uniform
external field. Proving long-range order in the Heisenberg ferromagnet
should be simpler than the problem of proving ferromagnetic long-range order
in an itinerant electron model such as the Hubbard model \cite{tasaki2003,froehlich2005}.
There are of course many other open problems in quantum spin systems. I will mention
some of them in the following sections. And many interesting problems have been solved
in the almost three decades since DLS1978. The ferromagnetic phase transition is a nagging
reminder that we lack the mathematical arguments to tackle some of the most basic
and clearly formulated problems of equilibrium statistical mechanics. 
Other problems in mathematical physics suffer from being physically not well-understood, from being
mathematically ill-posed, from being too complex, but not the Heisenberg ferromagnet.

In the years following DLS, Barry Simon made several other contributions to statistical mechanics
that will stand the test of time: elegant estimates, useful inequalities, and a great book 
\cite{aizenman1980,simon1980,simon1981a,simon1981b,simon1981c,simon1993}.
Is there a chance Barry will once more turn his attention to statistical mechanics?
The few years in the late seventies and the early eighties he worked in statistical mechanics
were a spectacular success. Maybe I can entice him with the following spectral problem for the Heisenberg ferromagnet which lies somewhere in the middle between his more recent interests 
and full-fledged statistical mechanics.

The problem is to prove that the Hamiltonian of the spin-1/2 isotropic Heisenberg ferromagnet on $\Ir^d$ 
in the GNS representation of one of its translation invariant pure ground states (which are
the fully polarized states) has purely absolutely continuous spectrum except for the ground state eigenvalue $0$. In one dimension this result follows from the work of Babbitt and Thomas
\cite{babbitt1977b}. So the open problem is for $d\geq 2$. To formulate the problem precisely
we introduce the relevant Hilbert space as $l^2(\F^d)$, where $\F^d$ is the set of finite
subsets of $\Ir^d$. Let $\{\Omega_X \mid X\in \F^d\}$ be the canonical orthonormal basis of $\H$.
One should think of $\Omega_X$ as the state vector with all spins pointing up except for those
finitely many that are located at $x\in X$, which are down.
The Hamiltonian $H$ is then the unique selfadjoint operator with domain containing the canonical
orthormal basis and such that
$$
H\Omega_\emptyset =0, \quad H\Omega_X=\sum_{x\in X, y\not\in X, \vert x-y\vert =1}
(\Omega_X - \Omega_{X\cup\{y\}\setminus\{x\}}) \,,\, \mbox{for all }  \emptyset\ne X\in\F^d\, .
$$
Note that, up to an overall sign, $H$ is also the generator of the symmetric simple
exclusion process on $\Ir^d$. Another expression for $H$ is given in \eq{xxx}.
It is easy to see that $H$ is non-negative definite and from the definition it is obvious that
$0$ is an eigenvalue. It is not hard to prove that the eigenvalue $0$ is simple. The conjecture
is that for all $d\geq 1$ the positive real line is absolutely continuous spectrum of $H$.
In particular, there are no strictly positive eigenvalues.

\section{General properties}

The mathematical setting for studying quantum spin systems rigorously,
as described in textbooks such as \cite{ruelle1969}, \cite{bratteli1997}, and \cite{simon1993},
was developed in the late sixties starting with Robinson \cite{robinson1967}. This is
not to say that there are no rigorous results about quantum spin models before 
that date. In particular Lieb and coworkers wrote several landmark papers that 
continue to be important today (see, e.g., \cite{lieb1961,lieb1962a}). In this general
formalism it was possible to obtain several fundamental theorems for large classes of
quantum spin models. An early example is the uniqueness of the equilibrium state
at any finite temperature for one-dimensional models with short-range interactions
by Araki \cite{araki1969}. In this section we first briefly review the main elements of the
general setup and then discuss some recent results in this context.

For the purpose of our discussion, a quantum spin is any quantum system with a 
finite-dimensional Hilbert space of states. A quantum spin system
consists of a finite or infinite number of spins, which we will label by  the elements  of a set
$V$. When $V$ is an infinite set, typically corresponding to the vertices of a lattice or a graph, 
one often considers families of  quantum spin systems, labeled by the finite subsets $X \subset
V$. Certain properties are easily stated for infinite sets $V$ directly, but to define the dynamics
and specific models we will first assume that $V$ is finite. In this case, the Hilbert space of states is
$$
\H_V=\bigotimes_{x\in V} \Cx^{n_x},
$$
where the dimensions $n_x\geq 2$ are related to the magnitude of the spins
by $n_x=2s_x+1$, and $s_x\in\{1/2,1,3/2,\ldots\}$. For each spin, the basic observables are the 
complex $n \times n$ matrices, which we will denote by $M_n$. The algebra of 
observables for the system is then
$$
\A_V=\bigotimes_{x\in V} M_{n_x}=\B(\H_V).
$$
Given a Hamiltonian, a self-adjoint observable $H_V = H_V^* \in \A_V$,
the Heisenberg dynamics of the system is defined as follows: for any $t \in \Rl$, 
an automorphism $\alpha_t$ is defined  on $\A$ by the formula
$$
\alpha^V_t(A)=U^*_V(t) A U_V(t)
$$
where $U_V(t)=e^{-itH_V}$.

One of the most important examples of a quantum spin system is the Heisenberg
model. In general, this model is defined on a graph $(V,E)$ which consists of a
set of vertices $V$ and a set of edges $E$. $E$ is a set of pairs of vertices denoted
by $e=(xy)$ for $x,y\in V$. Let $S^i_x$, $i=1,2,3$, denote the standard spin
$s_x$ matrices associated with the vertex $x$, and for each edge $e=(xy)$, let
$J_{xy} \in \Rl$ be a coupling constant corresponding to $e$. The Heisenberg
Hamiltonian (also called the XXX Hamiltonian) is then given by
\be
H_V=-\sum_{(xy)\in E} J_{xy} \vec{S}_x\cdot \vec{S}_y \, ,
\label{xxx}\ee
where $\vec{S}_x$ denotes the vector with components $S_x^1,S_x^2,S_x^3$.
The most commonly studied models are those defined on a lattice, such
as $\Ir^d$, with translation invariance. For such Hamiltonians, the
magnitude of the spins is usually constant, i.e., $s_x=s$, the edges
are the pairs $(xy)$ such that $\vert x-y\vert=1$, and  $J_{xy}
=J$. Depending of the sign of $J$, the Heisenberg model is said to 
be ferromagnetic ($J>0$) or the antiferromagnetic ($J<0$).

For extended systems, i.e., those corresponding to sets $V$ of infinite
cardinality, more care is needed in defining the quantities 
mentioned above. Hamiltonians are introduced as a sum of local terms
described by an interaction, a map $\Phi$ from the set of 
finite subsets of $V$ to $\A_V$, with the property that 
for each finite $X \subset V$, $\Phi(X) \in \A_X$ and $\Phi(X) = \Phi(X)^*$.
Given an interaction $\Phi$, the Hamiltonian is defined by
$$
H_V=\sum_{X\subset V} \Phi(X).
$$
Infinite systems are most often analyzed by considering families of 
finite systems, indexed by the subsets of $V$, and taking the
appropriate limits. For example, the $C^*$-algebra of 
observables, $\A$, is defined to be the norm completion of the 
union of the local observable algebras $\bigcup_{X\subset V}\A_X$.

To express the decay of interactions and correlation functions we need a
distance function on $V$. Let $V$ be equipped with a metric $d$. 
For typical examples, $V$ will be a graph and $d$ will
be chosen as the graph distance: $d(x,y)$ is the 
length of the shortest path (least number of edges)
connecting $x$ and $y$. The diameter, $D(X)$, of a finite subset 
$X \subset V$ is
$$
D(X)=\max \{ d(x,y)\mid x,y\in X\}.
$$

In order for the finite-volume dynamics to converge to a strongly continuous
one-parameter group of automorphisms on $\A$, one needs to impose
a decay condition on the interaction. For the sake of brevity, we will
merely introduce the norm on the interactions that will later appear
in the statement of the results described in detail here. For weaker conditions 
which ensure existence of the dynamics see 
\cite{bratteli1997,ruelle1969,simon1993,matsui1998}.
We will assume that the dimensions $n_x$ are bounded:
$$
N=\sup_{x\in V} n_x < \infty ,
$$
and that there exists a $\lambda >0$ such that the following
quantity is finite:
$$
\Vert \Phi \Vert_{\lambda} \, := \, \sup_{x \in V} \, \sum_{X \ni x}
\, |X|  \, \Vert \Phi(X) \Vert\, N^{2|X|} \, e^{ \lambda D(X)} < \infty.
$$
For translation invariant systems defined on a lattice with an interaction
$\Phi$ such that $\Vert \Phi\Vert_\lambda < \infty$ for some $\lambda >0$,
Lieb and Robinson \cite{lieb1972} proved 
a quasi-locality property of the dynamics, in the sense that, up to 
exponentially small corrections, there is a finite speed of propagation. 
More precisely, the obtained an estimate for commutators of the form
$$
\left[ \, \alpha_t(A) \, , \, B \, \right],
$$
where $t\in\Rl$, $A \in \A_X$, $B \in \A_Y$, and $X,Y\subset V$.
Clearly, such commutators vanish if $t=0$ and $X\cap Y=\emptyset$.
Quasi-locality, or finite group-velocity, as the property is also called,
means that the commutator remains small up to a time proportional
to the distance between $X$ and $Y$. We will now formulate
an extension of the Lieb-Robinson result to systems without
translation invariance or even an underlying lattice structure.

It will be useful to consider the following quantity
$$
C_B(x,t) :=  \sup_{A \in\A_{x}} \frac{ \| \left[ \, \alpha_t(A) \, , \,
    B \, \right] \|}{ \| A \|} \, ,
$$
for $x\in V$, $t\in \Rl$, $B\in \A_V$.
The basic result is the following theorem \cite{nachtergaele2005c}.

\begin{theorem}\label{thm:lr}
For $x\in V$, $t\in \Rl$, and $B\in\A_V$, we have the bound
\beann
C_B(x,t) &\leq& e^{2\, |t| \, \| \Phi \|_{\lambda}} C_B(x,0)\\
&&+  \sum_{y \in V: y \neq x} \,e^{- \, \lambda \,
  d(x,y)}  \left( e^{2 \, |t| \, \| \Phi \|_{\lambda}} - 1
\right)C_B(y,0) \, .
\eeann
\end{theorem}

Our proof avoids the use of the Fourier transform which seemed essential
in the work by Lieb and Robinson and appeared to be the main obstacle
to generalize the result to non-lattice $(V,d)$.

If the supports of $A$ and $B$ overlap, then the trivial bound $ \Vert
[\tau_t(A),B] \Vert \leq 2 \Vert A \Vert \Vert B \Vert$ is
better. Observe that for $B \in \A_Y$, one has that
 $C_B(y,0)\leq 2\Vert B \Vert \, \chi_Y(y)$, where
$\chi_Y$ is the characteristic function of $Y$, and therefore if $x
\not\in Y$, then one obtains for any $A \in \A_x$ a bound of the form
$$
\Vert [\tau_t(A),B]\Vert\leq 2\vert Y\vert \, \Vert A\Vert \Vert B\Vert 
\left(e^{2 \, |t| \, \| \Phi \|_{\lambda}} -1\right) e^{- \, \lambda
  \, d(x,Y)} \, , 
$$
where
$$
d(x,Y)=\min \{ d(x,y)\mid y\in Y\} \, .
$$
Moreover, for general local observables $A\in\A_X$, one may estmate
$$
\Vert [\tau_t(A),B]\Vert\leq N^{2\vert X\vert}\Vert A\Vert\sum_{x\in X} 
C_B(x,t) \, ,
$$
in which case, Theorem~\ref{thm:lr} provides a related bound.

Theorem \ref{thm:lr} has been further extended to provide bounds in the case
of  long-range interactions by Hastings and Koma \cite{hastings2005}.

The next general result relies on the Lieb-Robinson bound to prove
that a nonvanishing spectral gap above the ground state 
implies exponential decay of spatial correlations in the ground state. 
This result can be regarded as a non-relativistic analogue of the
Exponential Clustering Theorem in relativistic quantum field theory 
\cite{fredenhagen1985}. The idea that a Lieb-Robinson bound can be used 
as a replacement for strict locality in the relativistic context is natural but,
as far as I know, it was used in the literature for the first time only recently
by Hastings in \cite{hastings2004}.

In the physics literature the term {\em massive ground state} implies two
properties: a spectral gap above the ground state energy and exponential decay
of spatial correlations. It has long been believed that the first implies the
second, and the next theorem proves that this is indeed the case. The converse,
that exponential decay must be necessarily accompanied by a gap is not true in
general. Exceptions to the latter have been known for some time
\cite{nachtergaele1996}, and it is not
hard to imagine that a spectral gap can close without affecting the ground
state.

For simplicity of the presentation, we will restrict ourselves to the case where we have 
a representation of the system (e.g, the GNS representation) in which the model
has a unique ground state. This includes most cases with a spontaneously
broken discrete symmetry. Specifically, we will assume that our system
is represented on a Hilbert space $\H$, with a corresponding Hamiltonian $H \geq 0$, 
and that $\Omega \in \H$ is, up to a phase, the unique normalized vector state for which 
$H \Omega =0$.  We say that the system has a spectral gap if there exists $\delta >0$
such that $\spec (H) \cap (0,\delta) =\emptyset$, and in this case, 
the spectral gap, $\gamma$, is defined by
$$
\gamma=\sup\{\delta > 0 \mid \spec(H) \cap (0,\delta) =\emptyset\}.
$$

Our theorem on exponential clustering derives a bound 
for ground state correlations which take the form
\be
\langle\Omega, A\tau_{ib}(B)\Omega\rangle
\ee
where $b \geq 0$ and $A$ and $B$ are local observables. 
The case $b=0$ is the standard (equal-time) correlation
function. It is convenient to also assume a  
minimum site spacing among the vertices: 
\be
\inf_{ \stackrel{x,y \in V}{x \neq y}} d(x,y) =: a >0.
\ee
We proved the following theorem in \cite{nachtergaele2005c}.

\begin{theorem}[Exponential Clustering]\label{thm:decay}
There exists $\mu>0$ such that for any $x \neq y \in V$ and 
all $A\in\A_x$, $B\in\A_y$ for which
$\langle \Omega, B\Omega \rangle =0$, and $b$ sufficiently small,  
there is a constant $c(A,B)$ such that

\be
\left\vert \langle\Omega, A\alpha_{ib}(B)\Omega\rangle \right\vert
\leq c(A,B)e^{-\mu d(x,y)\left( 1 + \frac{\gamma^2 b^2}{4\mu^2d(x,y)^2}\right)}
\, .
\label{decay}\ee
One can choose 
\be
\mu = \frac{\gamma\lambda}{4 \Vert \Phi\Vert_\lambda + \gamma}\, ,
\ee
and the bound is valid for $0\leq \gamma b\leq 2\mu d(x,y)$.
\end{theorem}

The constant $c(A,B)$, which can also be made explicit, depends 
only on the norms of $A$ and $B$, (in its more general form) 
the size of their supports, and the system's minimum vertex spacing
$a$. For $b=0$, Theorem~\ref{thm:decay} may be restated as
\be
\left\vert \langle\Omega, AB\Omega\rangle
-  \langle\Omega, A\Omega\rangle\, \langle\Omega, B\Omega\rangle\right\vert 
\leq c(A,B) e^{-\mu d(x,y)} .
\label{zerob}\ee
Note that there is a trivial bound for large $b>0$
\be
\left\vert \langle\Omega, A\alpha_{ib}(B)\Omega\rangle \right\vert
\leq \Vert A\Vert \, \Vert B\Vert \, e^{-\gamma b} \, .
\ee
In the small $b>0$ regime, the estimate (\ref{decay}) can be viewed as 
a perturbation of (\ref{zerob}). Often, the important observation is
that the decay estimate (\ref{decay}) is uniform in the imaginary time $ib$,
for $b$ in some interval whose length, however, depends on
$d(x,y)$. 

In a recent work, Hastings and Koma have obtained an analogous result 
for models with long range interactions \cite{hastings2005}.

It has been known for some time that the spontaneous breaking of a 
continuous symmetry, such as the $SU(2)$ rotation symmetry of the XXX
Heisenberg model \eq{xxx}, precludes the existence of a spectral gap in the
infinite volume ground state. This is often called the Goldstone Theorem of
statistical mechanics \cite{landau1981, wreszinski1987}. Matsui \cite{matsui1997} adapted
the proof of the Goldstone theorem to show the absence of a spectral gap 
above the 111-interface ground states of the XXZ ferromagnet defined in \eq{xxz}
in all dimensions $d\geq 2$, a result anticipated by Koma and Nachtergaele
who had previously proved it for $d=2$ \cite{koma1996}.

In dimensions $\leq 2$ spontaneous continuous symmetry breaking cannot occur 
at positive temperatures.  This is called the  Mermin-Wagner-Hohenberg theorem
a general version of which was proved by Fr\"ohlich and Pfister \cite{froehlich1981}
and for systems of fractal dimension $d< 2$ by Koma and Tasaki \cite{koma1995}.
In the next section we review some results on existence of spontaneous 
symmetry breaking in quantum spin models at $T=0$ and $T>0$.

\section{N\'eel order and other applications of reflection positivity}

One of the main results in the 1978 paper by Dyson, Lieb, and Simon (DLS),
and an extension by Jordao Neves and Fernando Perez
\cite{jordao1986} to include ground states, and refinements
by Kennedy, Lieb, and Shastry \cite{kennedy1988a,kennedy1988b},
is the existence of long range order in the isotropic spin-$S$ antiferromagnet
on hypercubic lattice $\Ir^d$ at sufficiently small positive temperatures
in dimensions  $d\geq 3$, and any spin magnitude $S\geq 1/2$, and also
in the ground state ($T=0$), if $d\geq 2$ and $S\geq 1$, or $d=3$ and $S=1/2$.

DLS was a breakthrough based on another breakthrough, namely, the method
of infrared bounds to prove continuous symmetry breaking developed for classical
systems by Fr\"ohlich, Simon, and Spencer \cite{froehlich1976a,froehlich1976b}.
The method was soon generalized to cover a wide variety of models
\cite{froehlich1978b,froehlich1980} that have a property called reflection
positivity. There is no {\em a priori} reason why the infrared bound philosophy
should require reflection positivity, a property not shared by the 
Heisenberg ferromagnet \cite{speer1985}.
But in spite of serious attempts to find ways to prove infrared bounds
for quantum spin systems not relying on reflection positivity, no such
method has been found to date, and we still do not have a proof of long range
order in the ferromagnetic Heisenberg model at positive temperature.

For the XXZ model with Hamiltonian
\be
H_V=-\sum_{\vert x-y\vert=1} \frac{1}{\Delta}\left( S^1_x S^1_y
+S^2_x S^2_y\right) + S^3_x S^3_y\, , V\subset \Ir^d
\label{xxz}\ee
an Ising-type phase transition has been proved by Fr\"ohlich and Lieb 
\cite{froehlich1978b}, for $d\geq 2$ and sufficiently large $\Delta$ (since their proof assumed
reflection positivity). Kennedy \cite{kennedy1985} showed how to prove the phase
transition for all $\Delta >1$ without using reflection positivity but by instead relying on 
a version of the Peierls argument using a clever cluster expansion method. Thus it may
appear that to make further progress one should circumvent the reflection positivity property.
This is certainly true in part. It is worth mentioning however, that
the use of reflection positivity in quantum lattice models pioneered in DLS
has been successfully adapted to solve a number of other problems.
Let us just mention a few interesting examples: dimerization in the Hubbard-Peierls model 
on the ring and the spin-Peierls problem \cite{lieb1995a}, the flux-phase problem 
\cite{lieb1993,lieb1994}, and singlet ground states of quantum lattice systems 
\cite{lieb1989,freericks1995,lieb1999,lieb2000}

\section{The classical limit and beyond}

In a very elegant paper, \cite{lieb1973}, Lieb proved the existence of the classical limit
for a broad class of quantum spin systems for which the Hamiltonians can be 
considered as a multilinear function, $H$,  of the standard spin operators. The main result 
of Lieb's is then upper and lower bounds for the quantum partition function with the spin
operators taken to be of magnitude $S$. To state the result we define
$$
Z_Q(\beta,S)=\Tr e^{-\beta H(\{S_x^i\})}
$$
where the spin operators in the Hamiltonian are taken to be of magnitude $S$ and
the trace of the corresponding Hilbert space of dimension $(2S+1)^{\vert V\vert}$, and where
$V$ is fixed. The corresponding classical partition function is defined by
$$
Z_C(\beta) = \int e^{-\beta H(\{\Omega_x^i\})} \, \prod_{x\in V}d\Omega_x
$$
where $d\Omega_x$  is the normalized invariant measure on the 2-sphere (unit vectors
in $\Rl^3$). 
Then, Lieb proved
\be
Z_C(\beta) \leq Z_Q(\beta,S) \leq Z_C(\beta(1+\frac{c}{S})) 
\label{cls}\ee
with $c$ a universal constant. This result suffices to prove convergence of the mean free energy (assuming it exists for the classical system), as well as convergence of the expectation of averaged
quantities (intensive observabels).

In \cite{simon1980} Simon generalizes Lieb's result by proving the same 
relations for systems of $G$-spins,
where $G$ is an arbitrary compact Lie group. The Hamiltonian is assumed to be a multilinear
function operators $X_x^i$, which form the basis for the Lie-algebra of $G$ in a fundamental
representation of weight $\lambda$, copied at each $x\in V$. He then considers the sequence
of quantum systems where each $X_x^i$ is replaced by $\pi(X_x^i)$, with $\pi$ a representation
of weight $n\lambda$, $n=1,2,3,\ldots$. Simon's main result is then
\be
Z_C(\beta) \leq Z_Q(\beta,n) \leq Z_C(\beta(1+\frac{c}{n})) 
\label{cln}\ee
where $Z_C$ is again a classical partition function for a suitable classical system
which he defines.

The bounds \eq{cls} and \eq{cln} by themselves are not good enough to deduce
a phase transition for the quantum model at sufficiently large $S$ or $n$, assuming that the
classical model has a phase transition, although one may expect such a result.
Very recently, Biskup, Chayes, and Starr \cite{biskup2005} refined the Berezin-Lieb inequalities,
which are at the core of the results of Lieb and Simon, to include estimates of matrix elements
$e^{-\beta H}$ in coherent state vectors, not just for the partition function.
This enables them to prove the existence of phase transitions in the quantum model
for sufficiently large spin, assuming that both the classical and the quantum model are reflection 
positive and that chessboard estimates prove a phase transition for the
the classical model.  Another refinement of the classical limit, very much akin to studying the 
central limit theorem as a refinement of the law of large numbers, was undertaken in 
\cite{michoel2004,michoel2005}.

\section{Spin chains and Haldane's conjecture}

Quantum spin chains, i.e., systems defined on the one-dimensional lattice $\Ir$,
and their ground states are a fascinating subject by themselves. They have received a lot
of attention in recent years due to their relevance for quantum information theory.
But even before quantum information theory became again fashionable,
the subtleties and great variety in behavior attracted the attention of
condensed matter physicists and mathematical physicists alike. Haldane gave the 
subject
a big impetus when he predicted that the Heisenberg antiferromagnetic chain has 
qualitatively different ground states depending on whether the magnitude of the spins
is integer or not \cite{haldane1983}. 
The Bethe Asatz solution of the spin-1/2 chain shows a unique ground state
with polynomial decay of correlations and no spectral gap. Haldane conjectured that
the spin-1 chain would also have a unique ground state (in the thermodynamic limit),
but with a spectral gap above it and exponentially decaying correlations (this is sometimes
called the Haldane phase). 
Affleck and Lieb \cite{affleck1986} proved that for half-integer spin chains a unique
ground state indeed implies the existence of low-lying excitations.
A related implication was proved by Aizenman en Nachtergaele  \cite{aizenman1994}:
a unique ground state must necessarily have no faster than polynomial decay of correlations.

That the Haldane phase exists was rigorously demonstrated by Affleck, 
Kennedy, Lieb, and Tasaki, who proposed the spin-1 chain with 
Hamiltonian (now called the AKLT model)
\be
H_{[a,b]}=\sum_{x}\left[\frac{1}{3}
+\frac{1}{2} \vec{S}_x\cdot\vec{S}_{x+1}
+\frac{1}{6} ( \vec{S}_x\cdot\vec{S}_{x+1})^2\right]\quad.
\label{aklt}\ee
and proved  that it has a unique ground state, which they explicitly constructed, with 
exponential decay and a spectral gap above the ground state \cite{affleck1987}.

Inspired by the AKLT model and a construction by Accardi \cite{accardi1981}, 
Fannes, Nachtergaele, and Werner introduced Finitely Correlated States 
\cite{fannes1992} and proved that they provide the exact ground state of
a large family of spin chains, including the AKLT model, with similar properties.

All correlations of Finitely Correlated States (also called Quantum Markov States or Matrix Product States) can be explicitly computed by a transfer matrix type formula.
Let $\A_\Ir$ be the algebra of local observables of the spin chain with each
spin represented by a copy of $\A=M_n$, and let $\B$ be another finite-dimensional 
C$^*$-algebra, say $\B=M_k$.  Let $\E:\A\otimes\B\to\B$, be a
completely positive map that is unit preserving i.e., $\E(\idty_\A \otimes\idty_\B)=\idty_\B$.
It can be shown that for such $\E$ there exists at least one state $\rho$ on $\B$ such that 
$$
\rho(\E(\idty_\A\otimes B))=\rho(B), \mbox{ for all } B\in \B\quad. 
$$ 
Suppose $\rho$ is such a state. It is useful to introduce the maps
$\E_A:\B\to\B, \E_A(B)=\E(A\otimes B)$. It is then straightforward to verify that, with these
objects and properties, the following formula defines
a translation invariant state $\omega$ on $\A_\Ir$:
$$
\omega(A_1\otimes A_2 \otimes \cdots \otimes A_N)
=\Tr \rho \E_{A_1}\circ \E_{A_2}\circ \cdots\circ \E_{A_N}(\idty)\, .
$$
We say that $\omega$ is the finitely correlated state generated by
the triple $(\B,\E,\rho)$.

A particulary important class of maps $\E$ are the so-called {\em pure maps}.
Let $V:\Cx^k\to \Cx^d\otimes\Cx^k$ be linear and isometric, i.e., $V^*V=\idty$.
Then, $\E(A\otimes B)=V^*A\otimes BV$, defines a {\em pure} completely positive
unital map in the sense that there is no non-trivial decomposition
$\E=\E_1+\E_2$, with $\E_1$ and $\E_2$ completely positive maps. Such finitely
correlated states are called {\em purely generated}. Pure finitely correlated states
can be generated by a pure map $\E$ \cite{fannes1994}.

We conclude this section by specifying the triple $(\B,\E,\rho)$ that generates the ground state of the AKLT model with Hamiltonian \eq{aklt}. Since it is a spin-1 chain, $\A=M_3$. It turns out
that $\B= M_2$ and $\E(X)=V^* XV$, with $V:\Cx^2\to\Cx^3\otimes\Cx^2$,
given by
\beann
V\ket{1/2}&=&\sqrt{1/3}\ket{1}\otimes\ket{-1/2}
-\sqrt{2/3}\ket{0}\otimes\ket{1/2}\\
V\ket{-1/2}&=&\sqrt{1/ 3}\ket{-1}\otimes\ket{1/2}
-\sqrt{2/ 3}\ket{0}\otimes\ket{-1/2}
\eeann
where $\ket{m}$ denotes an eigenbasis  of $S^3$. $V$ is also the, up to a
phase factor, unique isometry satisfying the following intertwining relation
$$
V D^{(1/2)}=D^{(1)}\otimes D^{(1/2)} V
$$
where $D^{(S)}$ is the spin-$S$ representation of $SU(2)$.
The state $\rho$ is given by
$$
\rho(B)=\frac{1}{2}\Tr B\quad.
$$

\section{Perturbation Theory}\label{sec:perturb}

A major goal in perturbation theory of quantum spin systems is to show that
the set of interactions for which the model has a unique ground state with a
non-vanishing spectral gap above it (in the thermodynamic limit), is open in a
suitable topology on the space of interactions. Significant steps toward this
goal have been made by a number of authors and the results they obtained
have a scope and generality that justifies calling them quantum Pirogov-Sinai
theory \cite{datta1996,borgs1996,borgs1997} . It should be noted, however,
that the applicability of these results is restricted to models that can be regarded
as quantum perturbations of classical models.
In some examples suitable perturbation methods have also been used
to show the absence of gaps or eigenvalues \cite{datta2003,kennedy2005}.
The remarkable paper by Kennedy
and Tasaki \cite{kennedy1992} was the first to make a serious attempt
to  get away from perturbing classical models. A new result by Yarotsky, which
we discuss below, can be seen as taking that line of approach one step further. 

Yarotsky's result makes it possible to prove stability of the massive phase 
provided that there is a nearby Generalized Valence Bond Solid model
\cite{affleck1987,fannes1992,nachtergaele1996} that can be used as a reference 
point for the perturbation in the space of interactions. In particular, 
Yarotsky \cite{yarotsky2004} proves that the AKLT chain of \eq{aklt}
is contained in an open set of interactions with this property.

The general theorem proved by Yarotsky can be stated for
the following class of models defined on $\Ir^d$, $d\geq 1$.
For these models the Hilbert space $\H_x$, at $x\in\Ir^d$, 
is allowed to be infinite-dimensional. Let $\H_V=\bigotimes_{x\in V}
\H_x$, for any finite $V\subset\Ir^d$, be the Hilbert space associated
with $V$. The unperturbed model has finite-volume Hamiltonians
of the form
\be
H^0_V=\sum_{x, V_0 + x\subset V} h_x\, ,
\ee
where $h_x$ is a selfadjoint operator acting non-trivially only on
$\H_{V_0 + x}$, for some finite $V_0$.
The main assumption is then that there exists
$0\neq\Omega_x\in\H_x$ such that 
$\Omega^0_V=\bigotimes_{x\in V}\Omega_x$ is the unique
zero-energy ground state of $H^0_V$, with  a
spectral gap of magnitude at least $\vert V_0\vert$ above the
ground state. Explicitly:
\be
H^0_V \geq 0\, , \quad H^0_V\Omega^0_V=0\, ,\quad
H_V\ge 
\vert V_0\vert (\idty-\vert\Omega^0_V\rangle\langle\Omega^0_V\vert)\, .
\label{a0}\ee

The perturbed Hamiltonians are assumed to be of the form
\be
H_V = H^0_V + \sum_{x,V_0 + x \subset V}\phi_x^{(r)} + \phi_x^{(b)},
\label{a1}\ee
where $\phi_x^{(r)}$ and $\phi_x^{(b)}$ are selfadjoint operators
on $\H_{V_0 +x}$ satisfying
\be
\vert\langle \psi, \phi_x^{(r)}\psi\rangle|\le\alpha\|h_x^{1/2}\psi\|^2\, ,
\quad 
\|\phi^{(b)}_x\|\le \beta\, ,
\label{a2}\ee
for all $\psi \in \dom (h_x^{1/2})$, and suitable constants $\alpha$ 
and $\beta$.
One can call $\phi^{(r)}$ a ``purely relatively bounded'' perturbation, while
$\phi^{(b)}$ is simply a bounded perturbation.

\begin{theorem}[Yarotsky \cite{yarotsky2004}]\label{thm:perturb}
Let $H_V$ of the form \eq{a1}, satisfying assumption \eq{a0}.
For all $\kappa>1$ there exists $\delta=\delta(\kappa,d,V_0)>0$ 
such that if condition \eq{a2} is
satisfied with some $\alpha\in (0,1)$, and 
$\beta=\delta(1-\alpha)^{\kappa(d+1)}$, then\newline
1) $H_V$ has a non-degenerate gapped ground state
$\Omega_V:$ $H_V\Omega_V=E_V\Omega_V,$ 
and for some $\gamma>0$, independent of $V$, we have
\be
H_V\ge E_V \vert\Omega_V\rangle\langle\Omega_V\vert
+(E_V + \gamma)(\idty
-\vert\Omega_V\rangle\langle\Omega_V\vert)\, .
\ee
2) There exists a thermodynamic weak$^*$-limit of the ground
states $\Omega_V:$ 
\be
\langle
A\Omega_V,\Omega_V\rangle
\xrightarrow{V\nearrow\Ir^d} \omega(A),
\ee 
where $A$ is a bounded local observable.
\newline
3) There is an exponential decay of correlations in the infinite
volume ground state $\omega$: for some positive $c$ and $\mu$,
and $A_i\in\B(\H_{V_i})$,
\be
|\omega(A_1A_2)-\omega(A_1)\omega(A_2)|\le
c^{|V_1|+|V_2|} e^{-\mu d(V_1,V_2)}\|A_1\|\|A_2\|\, .
\ee
4) If, within the allowed range of perturbations, the terms $\phi_x$
(or the resolvents $(h_x+\phi_x-z)^{-1}$ in the case of unbounded
perturbations) depend analytically on some parameters, then the
ground state $\omega$ is also weak$^*$ analytic in these
parameters (i.e. for any local observable $A$ its expectation
$\omega(A)$ is analytic).
\end{theorem}

Application of this result to the AKLT model \eq{aklt} yields the following
theorem.

\begin{theorem}\label{thm:perturb_aklt}
Let $\Phi=\Phi^*\in\A_{[0,r]}$. Then there exists $\lambda_0>0$, such that 
for all $\lambda, |\lambda|<\lambda_0$, the spin chain with Hamiltonian
$$
H = H^{\rm AKLT} + \lambda \sum_{x} \Phi_x
$$
has a unique infinite-volume ground state with a spectral gap and
exponential decay of correlations.
Here $\Phi_x\in A_{[x,x+r]}$, is $\Phi$ translated by $x$.
\end{theorem}
 
To prove this theorem, Yarotsky shows that the AKLT model itself
can be regarded as a perturbation of a particular model, one he
explicitly constructs, to which Theorem \ref{thm:perturb} can be applied.

\section{Estimates of the spectral gap}

The most essential result in the AKLT paper \cite{affleck1987} is 
the proof of a non-vanishing spectral gap in the thermodynamic limit.
Also in \cite{fannes1992} one of the core results is an explicit lower
bound for the spectral gap of the Valence Bond Solid models, i.e., the
generalizations of the AKLT model introduced in that paper. A further
generalization to models with discrete symmetry breaking and hence a
finite number of distinct infinite-volume ground states, is studied in
\cite{nachtergaele1996}. In that paper lower bounds for the spectral
gap are obtained by an adaptation of the so-called martingale
method of Lu and Yau \cite{lu1993} to the quantum context.
In the following we will prove a simple version of this  method
for obtaining lower bounds for the spectral 
gap in extended systems. Then, we will discuss some generalizations and 
improvements. 

Consider a quantum spin model on $\A_\Ir$, with local Hamiltonians of the
form 
$$
H_{[a,b]}=\sum_{x=a}^{b-1} h_{x,x+1}
$$
where $h_{x,x+1}$ is the translation of a fixed nearest neighbor
interaction $h_{0,1}$, acting non-trivially only at the pair of sites $\{x,x+1\}$,
with $h_{0,1}$ a non-negative definite element of $\A_{[0,1]}$. Suppose that $\omega$ is 
a zero-energy ground state of this model in the sense that
$$
\omega(h_{x,x+1})=0,\ \mbox{for all } x\in \Ir.
$$
This is the situation encountered for VBS models and there are other 
interesting examples \cite{nachtergaele1996,contucci2002}.

Our aim is to show that, under some rather general conditions on $\omega$,
this model has a non-vanishing spectral gap above the ground state $\omega$,
i.e., there exists $\gamma>0$, such that
\be
\spec(H_\omega)\cap(0,\gamma)=\emptyset,
\label{gap_exists}\ee
where $H_\omega$ is the GNS Hamiltonian of the model in the state $\omega$,
which satisfies $H_\omega \Omega_\omega=0$.

The martingale method provides lower bounds for the smallest non-zero
eigenvalue of the finite-volume Hamiltonians $H_{[a,b]}$, uniform in 
the volume. We start with a lemma that implies that this is enough to
establish \eq{gap_exists}.

\begin{lemma}
Define 
$$
\gamma=\sup\{\delta > 0\mid \spec(H_\omega)\cap(0,\delta)=\emptyset\},
$$
assuming that the set on the RHS is non-empty and put $\gamma=0$
otherwise. Then
\be
\gamma\geq \liminf_{n\geq 2}\lambda_1(n),
\label{inf_vol_gap}\ee
where $\lambda_1(n)$ is the smallest non-zero eigenvalue of
$H_{[1,n]}$.
\end{lemma}
\begin{proof}
Let $\gamma_0$ be defined by the RHS of \eq{inf_vol_gap}. Then, we need
to show that
$$
\langle \Omega_\omega,\pi(X^*)H^3_\omega \pi(X) \Omega_\omega\rangle
\geq 
\gamma_0 \langle\Omega_\omega,\pi(X^*)H^2_\omega \pi(X) \Omega_\omega\rangle.
$$
for all $X\in\A_{\rm loc}$. 
We may assume that $X\in\A_{[a,b]}$, for some $a<b\in\Ir$. By using the
fact that $H_\omega\pi(X)\Omega_\omega \perp \ker H_\omega$, we see that
\beann
\langle \Omega_\omega,\pi(X^*)H^3_\omega \pi(X) \Omega_\omega\rangle
&=&\lim_{\Lambda_3\to\Ir}\lim_{\Lambda_2\to\Ir}\lim_{\Lambda_1\to\Ir}
\langle \Omega_\omega,\pi(X^*)\pi\left(
[H_{\Lambda_3},[H_{\Lambda_2},[H_{\Lambda_1},X]]] \right)\Omega_\omega\rangle\\
&=&
\langle \Omega_\omega,\pi(X^*) H_{[a-3,b+3]}^3\pi(X)\Omega_\omega\rangle.
\eeann
The last expression is the expectation of an element of $\A_{[a-3,b+3]}$
in $\omega$. Let $\rho_{[a-3,b+3]}$ be the density matrix of the restriction
of $\omega$ to $\A_{[a-3,b+3]}$. Then,
\beann
\langle \Omega_\omega,\pi(X^*)H^3_\omega \pi(X) \Omega_\omega\rangle
&\geq&
\Tr  \rho_{[a-3,b+3]} X^* H_{[a-3,b+3]}^3 X\\
&=& 
\Tr\left(H_{[a-3,b+3]}\rho_{[a-3,b+3]} X^* H_{[a-3,b+3]}\right)H_{[a-3,b+3]}\\
&\geq&
\lambda_1(b-a+5) \Tr \rho_{[a-3,b+3]} X^* H_{[a-3,b+3]}^2 X\\
&=&
\lambda_1(b-a+5) 
\langle \Omega_\omega,\pi(X^*)H^2_\omega \pi(X) \Omega_\omega\rangle
\eeann
As $X\in\A_\Lambda$ for all $\Lambda$ that contain $[a,b]$, we can take the
$\liminf$ of the last inequality to conclude the proof.
\end{proof}

With this lemma, the problem of finding a lower bound for the spectral
gap of the GNS Hamiltonian is reduced to estimating the smallest non-zero
eigenvalue of the local Hamiltonians. Due to the translation invariance
it is sufficient to consider intervals of the form $[1,L]$. As before,
\be
H_{[1,L]}=\sum_{x=1}^{L-1} h_{x,x+1}
\label{HL}
\ee
where $h_{x,x+1}$ is a translation of $h_{1,2}$, acting non-trivially
only at the nearest neighbour pair $\{x,x+1\}$, and we assume that
$h_{1,2}\geq 0$ and that $\ker H_{[1,L]} \neq\{0\}$. We will denote by
$\gamma_2$ the smallest nonzero eigenvalue of $h_{1,2}$, i.e., the gap
of $H_{[1,2]}$. It is obvious that
\be
\ker H_{[1,L]} =\bigcap_{x=1}^{L-1} \ker h_{x,x+1}
\label{kerH}
\ee
For an arbitrary
subset $\Lambda$ let $G_\Lambda$ be the orthogonal projection onto
\be
\ker \sum_{x, \{x,x+1\}\subset\Lambda} h_{x,x+1}
\label{GLambda}
\ee
For intervals $[a,b]$, $1\leq a < b\leq L$, $G_{[a,b]}$ is the
orthogonal projection onto the zero eigenvectors of $\sum_{x=a}^{b-1}
h_{x,x+1}$, and $G_{\{x\}}=\idty$ for all $x$. From these definitions
it immediately follows that the orthogonal projections $G_\Lambda$
satisfy the following properties:
\alpheqn\bea
G_{\Lambda_2}G_{\Lambda_1} &=& G_{\Lambda_1}G_{\Lambda_2}
= G_{\Lambda_2} \mbox{ if } \Lambda_1 \subset\Lambda_2
\label{GLambdaa}\\
G_{\Lambda_1}G_{\Lambda_2} &=& G_{\Lambda_2}G_{\Lambda_1}
\mbox{ if } \Lambda_1 \cap \Lambda_2 =\emptyset
\label{GLambdab}\\
h_{x,x+1} &\geq& \gamma_2(\idty-G_{[x,x+1]})
\label{GLambdac}
\eea\reseteqn
Define operators $E_n$, $1\leq n\leq L$, on $\H_{[1,L]}$ by
\be
E_n=\begin{cases} \idty-G_{[1,2]}       & \mbox{if } n=1\\
           G_{[1,n]}-G_{[1,n+1]} & \mbox{if } 2\leq n \leq L-1\\
            G_{[1,L]}         & \mbox{if } n=L \end{cases}
\label{defEn}\ee
One can then easily verify, using the properties
\eq{GLambdaa}-\eq{GLambdac}, that $\{E_n \mid1\leq n\leq L\}$
is a family of mutually orthogonal projections summing up to $\idty$,
i.e.:
\be
E_n^*=E_n,\qquad E_n  E_m=\delta_{m,n} E_n,
\qquad \sum_{n=1}^L E_n =\idty
\label{resolution}\ee

Next, we make a non-trivial assumption. 

\begin{assumption}\label{thm:ass}
There exists a constant $\epsilon$, $0\leq \epsilon
< 1/\sqrt{2}$, such that for all $1\leq n\leq L-1$
\be
E_n G_{[n,n+1]} E_n \leq \epsilon^2 E_n
\label{assa}\ee
or, equivalently,
\be
\Vert G_{[n,n+1]} E_n\Vert \leq \epsilon
\label{assb}\ee
\end{assumption}

Note that, due to \eq{GLambdaa}, $G_{[n,n+1]} E_n= G_{[n,n+1]}
G_{[1,n]} - G_{[1,n+1]}$. This relates Assumption \ref{thm:ass}
with Lemma 6.2 in \cite{fannes1992}, where an estimate for $\Vert G_{[n,n+1]}
G_{[1,n]} - G_{[1,n+1]}\Vert$ is given for general Valence Bond Solid
chains with a unique infinite volume ground state. The same
observation also implies that $[G_{[n,n+1]}, G_{[1,n]}]= [G_{[n,
n+1]}, E_n]$, which, if \eq{assb} holds, is bounded above in norm by
$2\epsilon$.

The next theorem is a special case of Theorem 2.1 in \cite{nachtergaele1996}. Just
like Theorem 6.4 in \cite{fannes1992} it provides a lower bound on the gap of
the finite volume Hamiltonians, but it achieves this in a slightly
more efficient way. We will repeat the proof for the particular case
stated  here, because it is simple, short, and instructive.

\begin{theorem}\label{thm:generalestimate}
With the definitions of above and under Assumption \ref{thm:ass}
the following estimate holds for the smallest non-zero eigenvalue
of $H_{[1,L]}$ :
\be
\lambda_1(L)
\geq \gamma_2 (1-\sqrt{2}\epsilon)^2
\label{gapestimate}\ee
i.e., the spectrum of $H_{[1,L]}$ has a gap of at least
$\gamma_2 (1-\sqrt{2}\epsilon)^2$ above the lowest eigenvalue, which is
$0$.
\end{theorem}

\begin{proof}
{}From the properties \eq{resolution} of the $E_n$ and the
assumption that $G_{[1,L]}\psi=0$, it immediately follows that
\be
\Vert\psi\Vert^2=\sum_{n=1}^{L-1}\Vert E_n\psi\Vert^2
\label{resolutionnormpsi}\ee
One can estimate $\Vert E_n\psi\Vert^2$ in terms of $\langle
\psi\mid h_{n,n+1}\psi\rangle$ as follows. First insert $G_{[n,n+1]}$
and the resolution $\{ E_m\}$:
\be
\Vert E_n\psi\Vert^2
=\langle\psi\mid(\idty-G_{[n,n+1]})E_n\psi\rangle
+ \langle\psi\mid \sum_{m=1}^{L-1} E_m G_{[n,n+1]}E_n\psi\rangle
\label{Enpsi}\ee
Using \eq{GLambdaa} and  \eq{GLambdab} one easily veryfies that $E_m$
commutes with $G_{n,n+1}$ if either $m\leq n-2$ or $m\geq n+1$. In
these cases $E_m G_{[n,n+1]}E_n= G_{[n,n+1]} E_m E_n=0$, because the
$E_n$ form an orthogonal family. By this observation we obtain the
following estimate. For any choice of constants $c_1,c_2>0$:
\bea
\Vert E_n\psi\Vert^2 &=&
\langle\psi\mid(\idty-G_{[n,n+1]})E_n\psi\rangle
+ \langle (E_{n-1}+E_n) \psi\mid G_{[n,n+1]}E_n\psi\rangle
\nonumber\\
&\leq& \frac{1}{2 c_1} \langle\psi\mid(\idty-G_{[n,n+1]})\psi\rangle +
\frac{c_1}{2}\langle\psi \mid E_n\psi\rangle
\label{Enpsi2}\\
&&\quad  +\frac{1}{2c_2} \langle\psi\mid E_n G_{[n,n+1]}
E_n\psi\rangle+\frac{c_2}{2}\langle\psi \mid (E_{n-1}+E_n)^2\psi\rangle
\nonumber\eea
where we have applied the inequality
$$
\vert\langle\phi_1\mid\phi_2
\rangle\vert\leq \frac{1}{2 c}\Vert \phi_1\Vert^2 +\frac{c}{2}\Vert\phi_2\Vert^2\quad ,
$$
for any $c>0$, to both terms of \eq{Enpsi}. The first term in the right
side of inequality \eq{Enpsi2} can be estimated with the interaction
using \eq{GLambdac}. The third term can be estimated with \eq{assa}.
It then follows that
$$
(2-c_1-\frac{\epsilon^2}{c_2})\Vert E_n\psi \Vert^2
-c_2 \Vert (E_{n-1}+E_n)\psi\Vert^2
\leq \frac{1}{c_1 \gamma_2}\langle\psi
\mid h_{n,n+1} \psi \rangle
$$
The term containing $E_{n-1}$ is absent for $n=1$, and
$E_n\psi=0$ if $n=L$.
We now sum over $n$ and use \eq{resolutionnormpsi} to obtain
$$
(2-c_1- \frac{\epsilon^2}{c_2}-2c_2)\Vert\psi\Vert^2
\leq \frac{1}{c_1 \gamma_{2}}\langle\psi
\mid H_{[1,L]} \psi \rangle
$$
Finally put $c_1=1-\epsilon\sqrt{2}$ and
$c_2=\epsilon/\sqrt{2}$ and
one obtains the estimate \eq{gapestimate} stated in the theorem.
\end{proof}

The theorem above, which is stated in the form given in \cite{nachtergaele1996}, can be
generalized in different directions. First,  one can formulate the conditions
for local Hamiltonians on a increasing sequence of finite volumes that are not
necessarily intervals, or that grow to infinity in a more restricted way.  
It is also not necessary that the model is translation invariant, as long as
the inequality of Assumption \ref{thm:ass} holds and $\gamma_2$ is  replaced
with the minimum of the gaps of $H_{[n,n+1]}$, over all $n=1,\ldots, L-1$. An
application where this is relevant was given in \cite{contucci2002}.

A nice improvement of the lower bounds provided by Theorem
\ref{thm:generalestimate} was made by Spitzer and Starr in \cite{spitzer2003}. We will
state their main result without proof. It is useful to generalize slightly, as
done in \cite{spitzer2003}, by introducing a family  $\epsilon_{m,n}$, for $m,n\geq 1$.
To this end, we consider finite systems on intervals of the form $[a,b]$, with
$a\leq -m , b\geq n$, and define $\epsilon(m,n)$, by
$$
\epsilon(m,n)=\sup\{\langle\psi,G_{[0,n]}\psi\rangle
\mid \psi\in\H_{[a,b]},\ G_{[-m,0]}\psi=\psi,\ G_{[-m,n]}\psi=0, 
\Vert\psi\Vert=1\}.
$$
Then, $\epsilon_{m,n}$, for $m,n\geq 1$, is defined by
$$
\epsilon_{m,n}=\sup_{m^\prime\geq m}\epsilon(m^\prime,n).
$$

\begin{theorem}[\cite{spitzer2003}]\label{thm:generalestimate-SpS}
With the same notations as in Theorem
\ref{thm:generalestimate}, we have the bound
$$
\lambda_1(L)\geq \lambda_1(n+m)\left(1-2\sqrt{\epsilon_{m,n}(1-\epsilon_{m,n})}
\right).
$$
\end{theorem}

Even in the case $m=n=1$, this theorem gives a strictly better lower bound
than Theorem \ref{thm:generalestimate}. To see this, note that
$\epsilon_{1,1}=\epsilon^2$, $\lambda_1(2)=\gamma_2$ and, for all 
$0\leq\epsilon\leq 1/\sqrt{2}$, one has
$$
1<\frac{1-2\sqrt{\epsilon^2(1-\epsilon^2)}}{(1-\sqrt{2}\epsilon)^2}\leq 2.
$$
So, depending on the value of $\epsilon$, the bound of Theorem 
\ref{thm:generalestimate-SpS} is always better than the one
provided by Theorem \ref{thm:generalestimate} by up to a factor of $2$.

By somewhat different but not unrelated methods excellent estimates
were obtained by Caputo and Martinelli for the
higher spin XXZ chains in \cite{caputo2003}.
They proved that the spectral gap is bounded above and below
by quantities of the form constant times $S$, supporting the
conjecture that the limit $\lim_{S\to \infty} \gamma(S)/S$ exists and is
strictly positive \cite{koma2001}.

\section{Ordering of Energy Levels}\label{sec:foel}

One easily checks that the Heisenberg Hamiltonian $H_V$, defined in
\eq{xxx}, commutes with both the total spin matrices and the 
Casimir operator given by
$$
S^i_V=\sum_{x\in V} S^i_x, \, i = 1, 2, 3, \quad \text{and} \quad
C=\vec{S}_V \cdot\vec{S}_V.
$$
The eigenvalues of $C$ are $S(S+1)$ where the parameter $S
\in \{ S_{\rm min}, S_{\rm min}+1,\ldots,S_{\rm max} \}$, with
$S_{\rm max}=\sum_x s_x$. $S$ 
is called the total spin, and it labels the irreducible 
representations of $SU(2)$. Let $\H^{(S)}$ be the eigenspace
corresponding to those vectors of total spin $S$. One can show that 
$\H^{(S)}$ is an invariant subspace for the Hamiltonian $H_V$ and
therefore, the number
$$
E(H_V,S) := \min \spec H_V \vert_{\H^{(S)}},
$$
is well-defined. 

Supported by partial results and some numerical calculations, we made the following conjecture
in \cite{nachtergaele2004}.

\begin{conjecture}[\cite{nachtergaele2004}]
All ferromagnetic Heisenberg models have the
{\em Ferromagnetic Ordering of Energy Levels} (FOEL) 
property, meaning
$$
E(H_V,S)< E(H_V,S^\prime), \text{ if } S^\prime < S.
$$
\end{conjecture}

\begin{figure}
\begin{center}
\resizebox{!}{13truecm}{\includegraphics{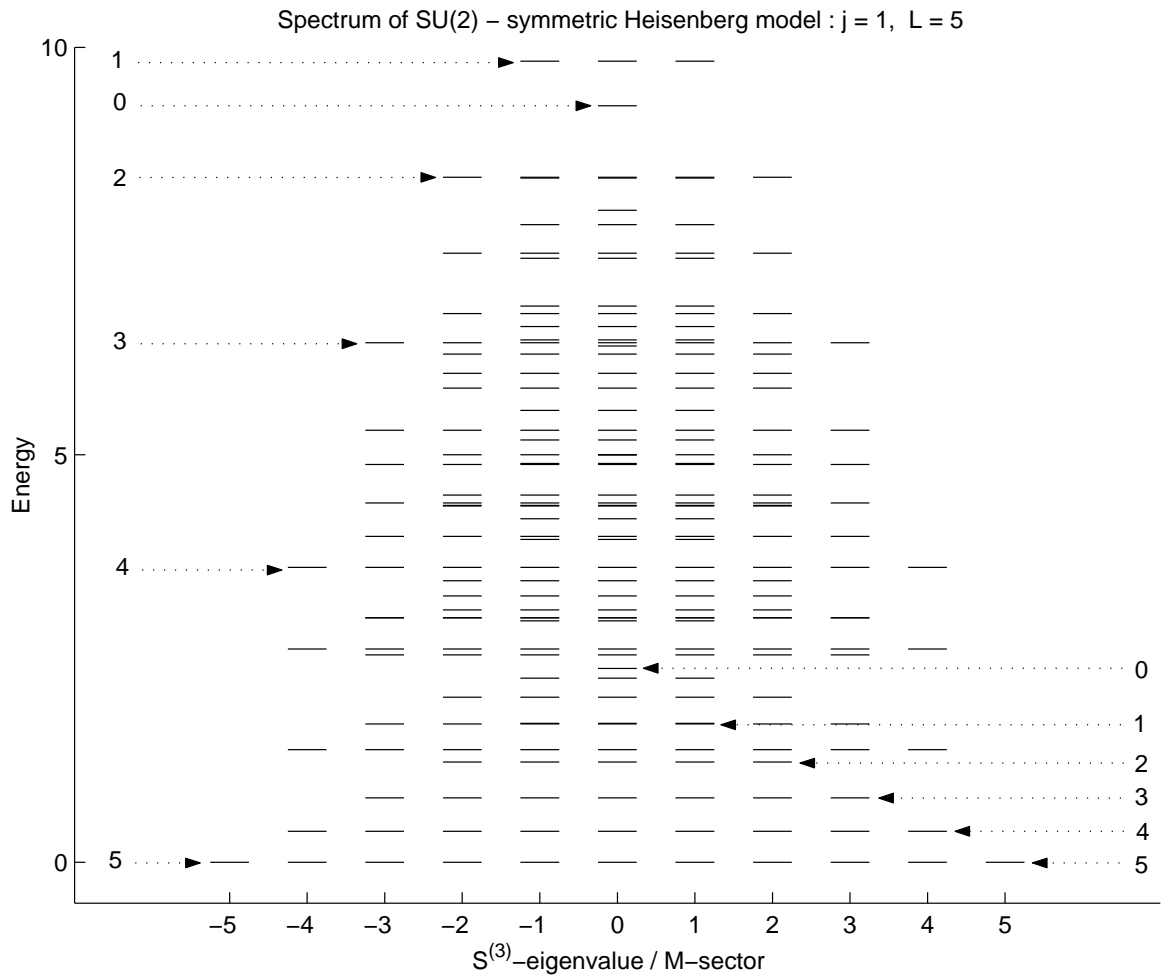}} 
\caption{\label{fig:foel}
The spectrum of a ferromagnetic Heisenberg chain  consisting of $5$ spin-$1$
spins with constant couplings.  On the horizontal axis we have plottted
the eigenvalue of the third component of the total spin. The spectrum is
off-set so that the ground state energy vanishes. The arrows on the right,
with  label $S$, indicate the multiplets of eigenvalues $E(H,S)$, i.e., the
smallest eigenvalue in the subspace of total spin $S$. The monotone ordering of
the spin labels is the FOEL property. On the left, we have indicated the
largest eigenvalues for each value of the total spin. The monotone ordering of
their labels in the  range $1,\ldots, 5$, is the content of the Lieb-Mattis
theorem \cite{lieb1962a} applied to this system.}
\end{center}
\end{figure}

In \cite{lieb1962a}, Lieb and Mattis proved ordering of energy levels 
for a class of Heisenberg models on bipartite graphs, which
includes the standard antiferromagnetic Heisenberg model. 
The FOEL property mentioned above can be considered as the 
ferromagnetic counterpart. To compare, a bipartite graph $G=(V,E)$ 
is a graph such that its set of vertices $V$ has a partition 
$V=A \cup B$ where $A \cap B = \emptyset$ and 
any edge $(xy)\in E$ satisfies either $x\in A$ and $y\in B$, or 
$x \in B$ and $y \in A$. For such a graph, one considers 
Hamiltonians of the form
$$
H=-H_V+H_A + H_B,
$$ 
where $H_V,H_A$, and $H_B$ are ferromagnetic Heisenberg Hamiltonians
on the graph $G$, and arbirary graphs $A$ and $B$, respectively.
Let  $S_X=\sum_{x\in X} s_x$, for $X=A,B, V$.
The Lieb-Mattis Theorem \cite{lieb1962a,lieb1989} then states that

(i) the ground state energy of $H$ is $E(H, \vert S_A - S_B\vert)$

(ii) if $\vert S_A - S_B\vert \leq S< S^\prime$, then $E(H, S) < E(H,S^\prime)$. 

One can see this property illustrated in Figure \ref{fig:foel}.

We first obtained a proof of FOEL for the spin 1/2 chain in 
\cite{nachtergaele2004}.
In that paper we also prove the same result for the ferromagnetic XXZ chain 
with $SU_q(2)$ symmetry. Later, in \cite{nachtergaele2005a}, we 
generalized the result to chains with arbitray values of the spin magnitudes
$s_x$ and coupling constants $J_{x,x+1}>0$. In short, we have the following
theorem.

\begin{theorem}
FOEL holds for all ferromagnetic chains.
\end{theorem}

The main tool in the proof is a special basis of SU(2) highest weight vectors introduced by
Temperley-Lieb \cite{Temperley1971} in the spin 1/2 case and by Frenkel and Khovanov 
\cite{frenkel1997} in the case of arbitrary spin. Some generalizations beyond 
the standard Heisenberg model have been announced  
\cite{nachtergaele2005a,nachtergaele2005b}.

The FOEL property has a number of interesting consequences.
The first immediate implication of FOEL is that the ground state energy of $H$
is $E(H,S_{\rm max})$, corresponding to the well-know fact that the ground state
space coincides with the subspace of maximal total spin.
Since there is only one multiplet of total spin $S_{\rm max}$, FOEL
also implies that the gap above the ground state is $E(H, S_{\rm max}-1)-E(H, S_{\rm max})$.
In the case of translation invariant models this is the physically expected property
 that the lowest excitations are simple spin-waves.

Another application of the FOEL property arises from the unitary equivalence of
the Heisenberg Hamiltonian and the generator of the Symmetric Simple Exclusion
Process (SSEP). To be precise, let $G=(V,E)$ be any finite graph, and define
$\Omega_n$  to be the configuration space of $n$ particles, for
$n=0,1,\ldots,\vert V\vert$, consisting of  $\eta:V\to \{0,1\}$, with 
$\sum_{x\in V} \eta(x)=n$. For any $(xy)\in E$, let $r_{xy}>0$. The SSEP is the
continuous time Markov process on $\Omega_n$ which exchanges the states
(whether there is a particle or not) at $x$ and $y$ with rate $r_{xy}$,
independently for each edge $(xy)$. The case $n=1$ is the random walk on $G$
with the given rates.

Alternatively, this process is defined by its generator on $l^2(\Omega_n)$:
$$
Lf(\eta) = \sum_{(xy)\in E}r_{xy}(f(\eta)-f(\eta^{xy})),
$$
where $\eta^{xy}$ is the configuration $\eta$ with the values at $x$ and $y$ 
interchanged. One verifies $L\geq 0$, $L1=0$, and therefore 
$L$ generates a Markov semigroup $\{e^{-tL}\}_{t\geq 0}$, such that
$$
\int f(\eta)\mu_t(d\eta)=\int (e^{-tL}f)(\eta) \mu_0(d\eta)\quad.
$$
where $\mu_0$ is the initial probability distribution on the particle 
configurations. It is easy to show that for each $n$ there is a unique
stationary measure given by the uniform distribution on $\Omega_n$. The
relaxation time, which determines the exponential rate of convergence to the
stationary state, is given by $1/\lambda(n)$, where $\lambda(n)>0$ is the
spectral gap (smallest eigenvalue $>0$) of $L$ as an operator on
$l^2(\Omega_n)$.

Aldous, based on discussions with Diaconis \cite{aldous}, made the following
remarkable conjecture concerning $\lambda(n)$:

\begin{conjecture}[Aldous]
$\lambda(n) = \lambda(1)$, for all $1\leq n\leq \vert V\vert -1$.
\end{conjecture}

Assuming the conjecture, one may determine the gap by solving the
one-particle problem.

To make the connection with FOEL, observe 
$$
\bigoplus_{n=0}^{\vert V\vert} l^2(\Omega_n) \cong (\Cx^2)^{\otimes\vert V \vert}\equiv \H_V
$$
This is the Hilbert space of a spin 1/2 model on $V$.
An explicit isomorphism is given by
$$
f\mapsto \psi=\sum_\eta f(\eta)\ket{\eta},
$$
where $\ket{\eta}\in\H$ is the tensor product basis vector defined by
$$
S^3_x\ket{\eta}=(\eta_x-1/2)\ket{\eta}
$$
Under this isomorphism the generator, $L$, of the SSEP becomes the
XXX Hamiltonian $H$. To see this it suffices to calculate the action
of $L$:
\beann
Lf(\eta)&\mapsto &\sum_\eta (Lf)(\eta)\ket{\eta}\\
&=&\sum_{(xy)}\sum_\eta r_{xy}(f(\eta)-f(\eta^{xy}))\ket{\eta}\\
&=&\sum_{(xy)} \sum_\eta r_{xy}(1-t_{xy}) f(\eta) \ket{\eta}\\
&=&H\psi
\eeann
where $t_{xy}$ is the unitary operator that interchanges 
the states at $x$ and $y$, and the last step uses
$$
1/2-2\vec{S}_x\cdot\vec{S}_y=1-t_{xy},
$$
and $J_{xy}=2r_{xy}$.

The number of particles, $n$, is a conserved quantity for SSEP. Since,
$S^3_{\rm tot} = -\vert V \vert/2 +n$, the corresponding conserved quantity for
the Heisenberg model is the third component of the total spin. Under the
isomorphism the invariant subspace of all functions $f$  supported on
$n$-particle configurations is identified with the set of vectors with
$S^3=S^3_{\rm max} -n$, which we will denote by $\H_n$. The unique invariant
measure of SSEP for $n$ particles on $V$,  which is the
uniform measure, corresponds to the ferromagnetic ground state with
magnetization $n -\vert V \vert/2$. The eigenvalue $\lambda(n)$ is the
spectral gap of $H_V\vert_{\H_n}$. It is then easy to see that the
FOEL property implies that $\lambda(n)=\lambda(1)$.  Since we proved FOEL
for chains, we also provided a new proof of Aldous' Conjecture in that case
\cite{handjani1996}. 

\section{Other topics}\label{sec:other}

To conclude I would like to give a sample of the many topics in quantum spin systems
that I have omitted in the preceding sections. First,  I have not discussed 
developments in 
exact solutions of quantum spin chains by the Bethe ansatz and related techniques,
which is a subject all by itself. I also did not elaborate on the many recent results on the XXZ
model related to interface and droplet states. For a short review on this topic see
\cite{nachtergaele2001b}.
There are also some interesting connections  between quantum spin chains and
problems in combinatorics related to counting alternating sign matrices and
packed loop configurations on a finite square lattice with matchings specified 
at the boundary \cite{razumov2001a,razumov2001b,degier2005}.
Another topic of recent activity is the behavior of the dynamics and non-equilibrium 
properties of quantum spin chains \cite{matsui2003,nachtergaele2003,ogata2002,ogata2004}.

Quantum spin systems are an inexhaustible source of mathematical problems, many of which are of
physical interest. Somebody should write a book about them.

\bigskip
\noindent
{\em Acknowledgements.} 
This article is based on  work supported by the U.S. National Science Foundation under Grant \# DMS-0303316.
I have freely borrowed and modified some passages from my recent preprints 
\cite{nachtergaele2005b} (with Shannon Starr) and \cite{nachtergaele2005e}
(with Robert Sims), and my unpublished lecture notes of the Volterra-CIRM International School on
{\em Quantum Markov chains and their applications
in physics and quantum information},Trento, Italy, 14-20 December 2001.
I also thank Robert Sims and an anonymous referee for pointing several errors
in a previous draft.

\providecommand{\bysame}{\leavevmode\hbox to3em{\hrulefill}\thinspace}

\end{document}